\documentclass[a4paper,pra,showpacs,superscriptaddress,twocolumn]{iopconfs}
\usepackage{amsfonts}
\usepackage{amsmath,amssymb,latexsym}
\usepackage{calc}
\usepackage{graphicx}
\usepackage{bm}

\begin{document}

\title{Homogeneous  linewidth  of  the  $^4$I$_{11/2}$--$^4$I$_{15/2}$
  optical transition of erbium in LiNbO$_3$:Er$^{3+}$}

\author{G. Mandula, Z. Kis, P. Sinkovicz, and L. Kov\'acs}
\address{Research Institute for Solid State Physics and Optics\\
  1121 Budapest, Konkoly-T. Mikl\'os \'ut 29-33, Hungary}

\ead{zsolt@szfki.hu}

\begin{abstract}
  We  work  out a  simple,  pulsed  pump-probe  measurement scheme  to
  measure  the homogeneous  linewidth of  an atomic  transition  in an
  inhomogeneously   broadened   spectral  line   in   a  solid   state
  environment.   We   apply  the  theory  to   the  $^4$I$_{11/2}$  --
  $^4$I$_{15/2}$ optical  transition of erbium  in LiNbO$_3$:Er$^{3+}$
  crystal.   Beside  obtaining  the  homogeneous  linewidth,  we  have
  estimated the population relaxation time as well.
\end{abstract}
\date{\today}

\section{Introduction}

The  successful  realization of  coherent  quantum control  procedures
imposes serious  requirements on  the participating fields  and atomic
systems. The most  important ones are the following:  the light fields
should be sufficiently coherent,  the population- $T_1$ and the dipole
relaxation time $T_2$ of the  involved atomic quantum states should be
sufficiently  long. Experience shows  that in  a solid  the electronic
spins  are  the most  robust  quantum  objects.   Some of  the  recent
achievements and  feasible proposals in the field  of coherent control
in  solids  are  electromagnetically induced  transparency  \cite{Ham,
  Ichimura98}  (exp);  coherent  Raman beats  \cite{Louchet07}  (exp);
coherent  population  trapping in  ruby  crystal  at room  temperature
\cite{Kolesov05}  (exp);  slow  down   of  light  pulses  in  a  solid
\cite{Turukhin02,Kuznetsova02}  (exp,   theor)  \cite{Sun05}  (theor);
quantum  storage of  light  pulses \cite{Turukhin02,Longdell05}  (exp)
\cite{Moiseev03} (theor).  Coherent  population transfer to the doping
atoms  has  also been  performed  experimentally \cite{Goto07},  which
opens the way for the realization of quantum logic gates.

In a  solid the $T_2$ time  is often orders of  magnitude shorter than
the $T_1$  time, hence  it is the  main limiting factor  for realizing
coherent control experiments. The  standard way of measuring the $T_2$
time  is  the photon  echo spectroscopy  \cite{Stenholm}. Photon  echo
measurements  have been  done  for  more than  forty  years in  solids
\cite{Kurnit64, Abella66, Macfarlane81, Equall95, Guillot-Noel07}.

In this paper we propose an alternative technique to measure the $T_2$
dipole  relaxation time  based on  spectral hole  burning spectroscopy
\cite{Stenholm, Macfarlane02}.   Our method is applicable  if the atom
can be  treated as  an effective two-level  system interacting  with a
narrow  linewidth  coherent  laser field.   It  is  shown  that  for  a
sufficiently low  intensity pump  and probe fields,  the width  of the
recorded  absorption line-profile is  proportional to  the homogeneous
linewidth  of  the  spectral  line.   The method  is  applied  to  the
transition   $^4$I$_{11/2}$    --   $^4$I$_{15/2}$   of    erbium   in
LiNbO$_3$:Er$^{3+}$.   A   very  detailed  fluorescence  spectroscopic
analysis  of  the transition  lines  of  LiNbO$_3$:Er$^{3+}$ has  been
reported in  \cite{Gill96}.  The photon echo measurement  of the $T_2$
dipole  relaxation  time   of  the  $^4$I$_{13/2}$  --  $^4$I$_{15/2}$
transition has been published in \cite{Sun02}.

The organization  of the paper  is as follows: in  section \ref{model}
the  model  system is  presented.   The  pulsed  pump-probe scheme  is
described and  the measurable  absorption-line profile is  derived. In
section \ref{experiment} the details of the experiment and the results
of  the measurement  are  presented.  The  results  are summarized  in
section \ref{summary}.

\section{Model calculation}
\label{model}
We consider a set of effective two-state atoms interacting with a long
pump-  and  a short,  read-out  probe  pulse.   The atomic  transition
frequencies  $\omega_{eg}$  are spread  in  a  range  ${\cal D}$,  the
distribution  of the  atoms over  this range  is characterized  by the
function $g(\omega_{eg})$, where the indices  $g$ and $e$ refer to the
atomic  ground  and excited  states,  respectively.  The  distribution
$g(\omega_{eg})$  describes   an  inhomogeneous  line   broadening  in
spectroscopic language.  We use  the Bloch vector notation to describe
the quantum state of the two-state atoms.  The components of the Bloch
vector   $[U_{\delta}(t),  V_{\delta}(t),  W_{\delta}(t)]^T_{\delta\in
  {\cal D}}$  can be expressed in  terms of the  atomic density matrix
elements.  They are defined as
\begin{subequations}
\begin{eqnarray}
U_{\delta}(t)&=&\varrho_{eg}(\delta; t)+\varrho_{ge}(\delta; t)\,,
\\
V_{\delta}(t)&=&i[\varrho_{eg}(\delta; t)-\varrho_{ge}(\delta; t)]\,,
\\
W_{\delta}(t)&=&\varrho_{ee}(\delta; t)-\varrho_{gg}(\delta; t)\,,
\end{eqnarray}
\end{subequations}
where the index $\delta$  denotes the frequency difference between the
transition  frequency $\omega_{eg}$  of the  particular atom  and some
reference   frequency   $\omega$,   $\delta=\omega_{eg}-\omega$.

Now we consider the interaction  between the atoms and a monochromatic
light wave.  The dynamics is described by the Bloch equations
\begin{subequations}
\label{UVW equation}
\begin{eqnarray}
  \dot{U}_{\delta} &=& - \delta V_{\delta} - \frac{1}{T_2} U_{\delta}\,,
  \\
  \dot{V}_{\delta} &=& \Omega_{\rm w} W_{\delta} + \delta U_{\delta} - \frac{1}{T_2} V_{\delta}\,,
  \\
  \dot{W}_{\delta} &=& - \Omega_{\rm w} V_{\delta} - \frac{1}{T_1} \left( W_{\delta}+1 \right)\,,
\end{eqnarray}
\end{subequations}
where $T_1$  ($T_2$) denotes the population  (dipole) relaxation time.
The  reference frequency  is equal  to  the angular  frequency of  the
field, $\omega=\omega_{\rm w}$.  The Rabi frequency $\Omega_{\rm w}$ is
given  by $\Omega_{\rm  w} =  -d_{eg}E_{\rm w}/\hbar$,  where $d_{eg}$
denotes  the  dipole moment  of  the  atomic  transition, $E_{\rm  w}$
describes the pump (write $\equiv$  w) field strength.  The pump field
acts so long on the  atoms that the transitions become saturated, i.e.
the  atoms achieve  a  steady  state.  The  steady  state solution  to
Eqs.~(\ref{UVW equation}) is given by
\begin{subequations}
\label{UVW solution}
\begin{eqnarray}
 U_{\delta}(0) &=& \frac{\Omega_{\rm w} \delta}{\delta ^2 + \frac{1}{T_2 ^2} + \frac{T_1}{T_2} \Omega_{\rm w} ^2}\,,
\\
 V_{\delta}(0) &=& - \frac{\Omega_{\rm w} / T_2}{\delta ^2 + \frac{1}{T_2 ^2} + \frac{T_1}{T_2} \Omega_{\rm w} ^2}\,,
\\
 W_{\delta}(0) &=& - \frac{\delta^2 + 1 / T_2 ^2}{\delta ^2 + \frac{1}{T_2 ^2} + \frac{T_1}{T_2} \Omega_{\rm w} ^2}\,.
\end{eqnarray}
\end{subequations}
The origin of time is set to the end of the pumping process. Hence the
pumping process  starts at a time  instant $t\ll0$ and  ends at $t=0$.
The second, probe pulse arrives after a time delay $t_d$ that the pump
pulse ceases. Assuming  that $T_2\ll t_d$ the components  of the Bloch
vector are $U_{\delta}(t_d)=0$, $V_{\delta}(t_d)=0$, and
\begin{equation}
  W_{\delta}(t_d)=(1+W_{\delta}(0))e^{-t_d/T_1}-1\,.
\end{equation}
In rare-earth doped  LiNbO$_3$ it is generally true  that $T_2\ll T_1$
for the ionic decay times.  It is assumed that the length of the probe
pulse  $\tau$   satisfies  the  relation   $T_2\ll\tau\ll  T_1$.   The
interaction  between  the probe  pulse  and  the  two-state atoms  are
described   again  by  Eqs.~(\ref{UVW   equation}),  where   the  Rabi
frequencies $\Omega_{\rm w}$ should  be replaced by the Rabi frequency
associated   with  the   probe  pulse   $\Omega_{\rm  p}=-d_{eg}E_{\rm
  p}/\hbar$.  If  the above conditions  hold for the time  scales, and
$\tau\Omega_{\rm    p}\ll1$,   then    $W_{\delta}(t_d+\tau)   \approx
W_{\delta}(t_d)$, i.e. the  inversion is approximately constant during
the action of  the probe pulse.  As a result,  for the coherences one
obtains
\begin{subequations}
\label{UVW solution 2}
\begin{eqnarray}
 U_{\delta}(t_d+\tau) &=& -\frac{(\Delta-\delta)\Omega_p}{(\Delta-\delta) ^2 + \frac{1}{T_2 ^2}}
W_{\delta}(t_d)\,,
\\
 V_{\delta}(t_d+\tau) &=&  \frac{\Omega_p/T_2}{(\Delta-\delta) ^2 + \frac{1}{T_2 ^2}}
W_{\delta}(t_d)\,,
\end{eqnarray}
\end{subequations}
where $\Delta$ is  the detuning of the probe  pulse from the frequency
of the pump  (write) pulse. The susceptibility of  the ensemble of the
atoms     is      obtained     from     $U_{\delta}(t_d+\tau)$     and
$V_{\delta}(t_d+\tau)$  by summing  up for  all subset  of  atoms with
transition frequency difference $\delta$. The sum should be weighted by the distribution function $g(\delta)$

\begin{figure}
\begin{center}
\includegraphics[width=12cm]{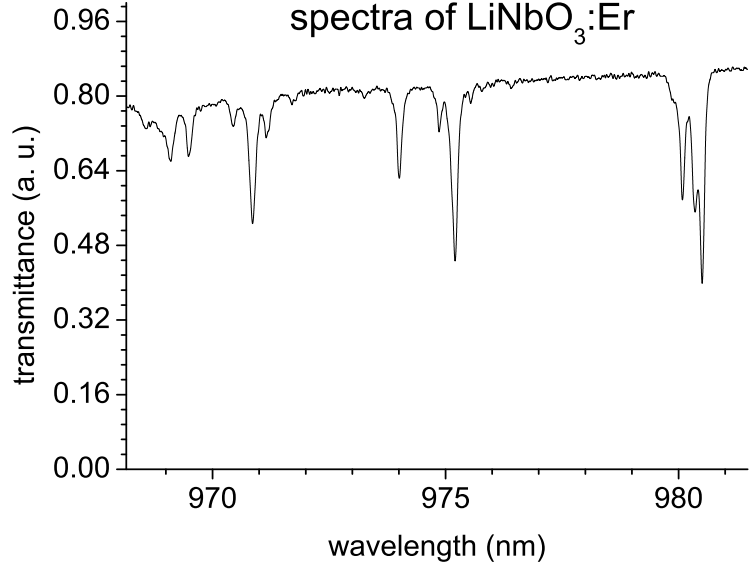}
\end{center}
\caption{A section of  the absorption spectrum of LiNbO$_3$:Er$^{3+}$.}
\label{fig1}
\end{figure}

\begin{equation}
  \label{chidef}
  \chi^+(\Delta)=-{\cal N}\frac{|d_{eg}|^2}{\varepsilon_0\hbar}
  \int g(\delta)\frac{(\Delta-\delta)+i/T_2}{(\Delta-\delta) ^2 
    + \frac{1}{T_2 ^2}} W_\delta(t_d)\,d\delta \,,
\end{equation}
where ${\cal N}$  is the density of the  two-state atoms participating
in  the process.   For a  wide enough  distribution  $g(\delta)$, i.e.
${\cal  D}\gg T_2^{-1}$, the  function $g(\delta)$  can be  brought in
front of the integral.  As a result, the Eq.~(\ref{chidef}) simplifies
to  a convolution  of a  Lorentzian-curve and  $W_{\delta}(t_d)$.  The
convolution  can  be  performed  by  the application  of  Fourier  and
inverse-Fourier transforms. The results is given by
\begin{eqnarray}
  \chi^+(\Delta)=\frac{{\cal N}|d_{eg}|^2}{\varepsilon_0\hbar}\pi g(0)
  \left[
    \frac{\frac{T_1}{T_2}\Omega_{\rm w}^2\Delta}
    {\Gamma\left((\frac{1}{T_2}+\Gamma)^2+\Delta^2\right)}
    +i\left(
      1-\frac{\frac{T_1}{T_2}\Omega_{\rm w}^2(\frac{1}{T_2}+\Gamma)}
    {\Gamma\left((\frac{1}{T_2}+\Gamma)^2+\Delta^2\right)}e^{-t_d/T_1}
    \right)
  \right]\,,
\label{convolve}
\end{eqnarray}
where $\Gamma^2=\frac{1}{T_2^2}+\frac{T_1}{T_2}\Omega_{\rm w}^2$.  The
real part  of $\chi^+(\Delta)$ describes  a phase shift for  the probe
pulse,  whereas the  imaginary  part contributes  to attenuation.  For
$\Delta=0$,  i.e.   when the  pump  and  probe  pulses have  the  same
frequency, the attenuation is  minimal.  Therefore, the imaginary part
of  the  susceptibility   in  Eq.~(\ref{convolve})  describes  a  {\em
  spectral hole}.

In the  Eq.~(\ref{convolve}) the quantity  $(T_1/T_2)\Omega_{\rm w}^2$
corresponds  to power broadening.   For a  low intensity  pump (write)
pulse the result of the convolution can be expanded, in lowest order of
$\Omega_{\rm w}^2$ one finds
\begin{equation}
  \label{line-shape}
  \textrm{Im}(\chi^+(\Delta))=\frac{{\cal N}|d_{eg}|^2}{\varepsilon_0\hbar}
  \pi g(0)\left(1-\frac{2\frac{T_1}{T_2}\Omega_{\rm w}^2}
    {\Delta^2+\frac{4}{T_2^2}}e^{-t_d/T_1}\right)\,.
\end{equation}
The exponential term describes the  vanishing of the spectral hole due
to population  relaxation. Comparing  the $\Delta$ dependence  of this
equation         with         a        Lorentzian         distribution
$\approx[\Delta^2+(\sigma/2)^2]^{-1}$,  one observes that  by scanning
the probe  field frequency  through resonance with  the pump  field, a
Lorentzian line-shape is obtained with half-width of $\sigma= 4T_2^{-1}$.

\section{Experiment}
\label{experiment}

\begin{figure}
\begin{center}
\includegraphics[width=15cm]{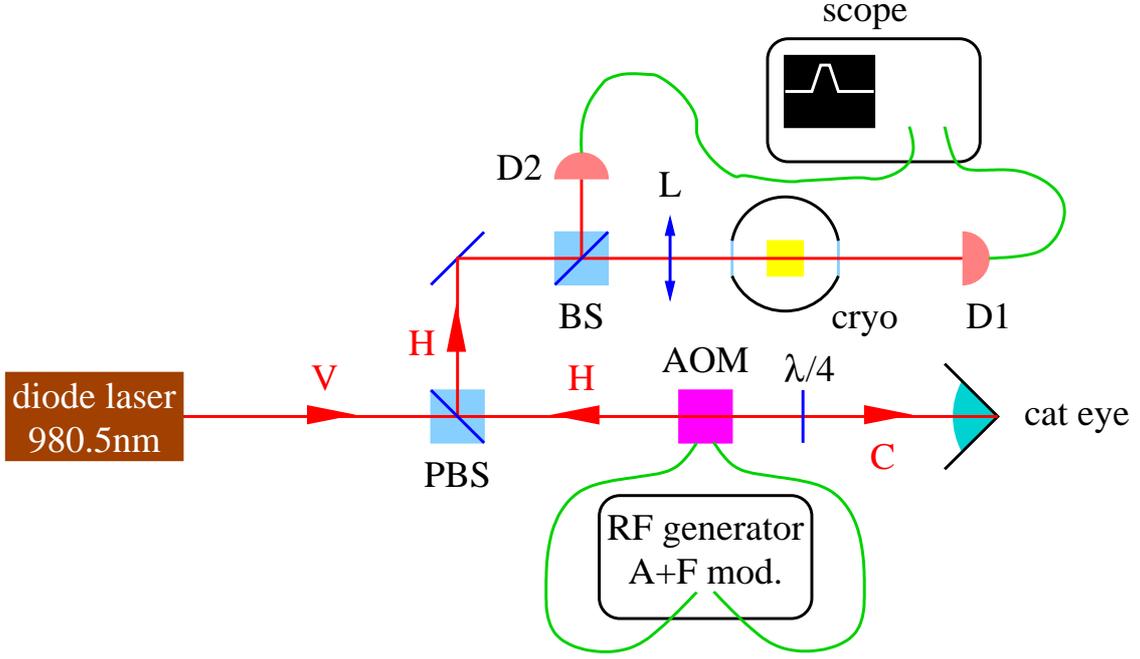}
\end{center}
\caption{The experimental setup for the $T_2$ measurement. The symbols
  mean: BS --  beam splitter; PBS -- polarizing  beam splitter; AOM --
  acousto  optic  modulator;  $\lambda/4$  --  retarder  plate;  L  --
  focusing lense; cryo -- cryostat with lowest temperature of 8.2K.}
\label{fig2}
\end{figure}

We  have realized  the pump-probe  process described  in  the previous
section  for erbium  doped LiNbO$_3$  to measure  the  $T_2$ dipole
relaxation  time  for  the   optical  transition  between  the  states
$^4$I$_{11/2}$--$^4$I$_{15/2}$ of  Er$^{3+}$.  The absorption spectrum
of a stoichiometric LiNbO$_3$:Er$^{3+}$ is shown in figure \ref{fig1}.
The   group    of   lines   around   980.5nm    corresponds   to   the
$^4$I$_{11/2}$--$^4$I$_{15/2}$  transition.   The  spectral lines  are
incoherently broadened, the with  of the deepest peak is approximately
0.8cm$^{-1}\equiv$  24~GHz. The  scheme of  our experimental  setup is
shown in figure \ref{fig2}. An external cavity diode laser is tuned to
the center  of the peak  at 980.5nm. The  spectral width of  the laser
line is approximately 1MHz. The pump and probe pulses are derived from
the  same  seed,  the  amplitude  is modulated  by  an  acousto  optic
modulator (AOM).  In order to  correct the walk-off of  the diffracted
beam after the AOM, a cat eye configuration is applied.

The pump pulse has a fixed frequency, whereas the frequency of the probe pulse
should be scanned  through the pump frequency.  First we  made the scanning by
varying the radio frequency on the  AOM. It turned out that the scanning range
$\approx$~50MHz  of the  AOM is  not sufficient,  hence we  used the  built in
scanning  capability of  the diode  laser with  a frequency  range  of several
GHz. The light beam leaving the cat eye system is fed into the crystal sample,
which is placed in a cryostat  with lowest temperature of 8.2K. The signal and
the  reference  beam  intensity is  measured  by  the  D1 and  D2  photo-diode
detectors,   respectively.   The  sample   stoichiometric  LiNbO$_3$:Er$^{3+}$
crystal has  been grown  in-house by top-seeded  sulution growth  method.  The
concentration of Er in the solution from which the crystal has been grown, was
0.1 mol\%.

\begin{figure}
\begin{center}
\includegraphics[width=12cm]{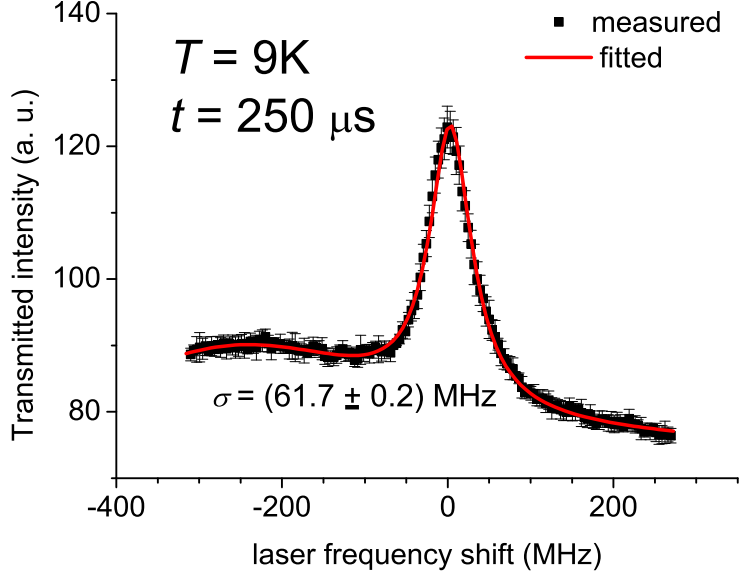}
\end{center}
\caption{The  result of several  scans with  the probe  pulse through
  resonance with the pump pulse. }
\label{fig3}
\end{figure}

The total  measurement cycle consists of  three stages: a  long pumping period
with duration more  than $4500\,\mu$s, a delay $t_d$  of more than $100\,\mu$s
when the light field is switched off,  and a scanning with a probe field which
takes $200\,\mu$s.  The delay time is measured from the switch off of the pump
field until the middle  of the scan period. The time delay  and the scan times
are limited by the piezo-scanning rate  of the diode laser head.  On the other
hand, the susceptibility in Eq.~(\ref{convolve})  has been derived for a fixed
frequency probe pulse.  Hence the rate of change of  the probe field frequency
should be small  enough so that the system can relax  with the rate $T_2^{-1}$
to the quasi steady state determined  by the probe pulse.  We have verified by
numerically  solving the  dynamical Eqs.~(\ref{UVW  equation}) that  using the
above  parameters the  system relaxes  to the  quasi steady  state  for slowly
varying probe  field detuning $\Delta$.  We  have used the  measured $T_1$ and
$T_2$ times  in the simulations.  Details  of the $T_1$  measurement are given
later in this section.  In our experiment the scans have been repeated several
times to reduce the measurement errors.  The average of several scans is shown
in figure \ref{fig3} for $t_d=250 \mu$s and $T=9$K.

\begin{figure}
\begin{center}
\includegraphics[width=12cm]{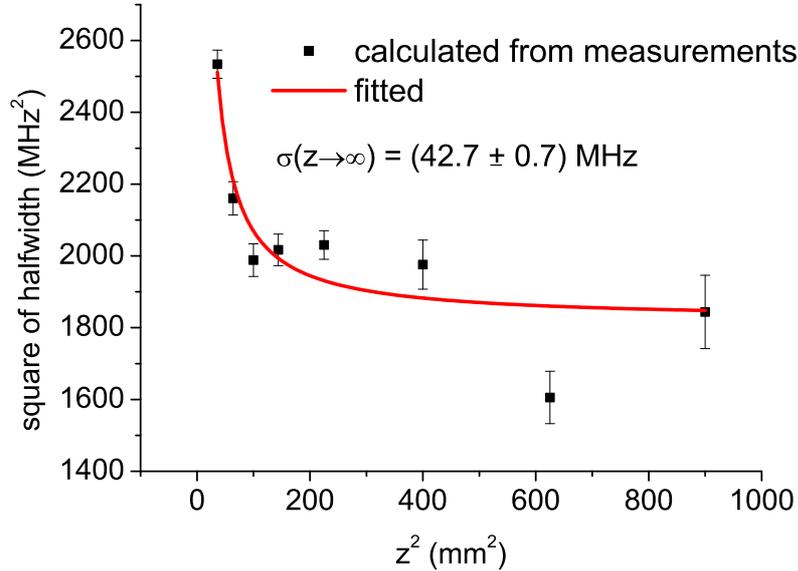}
\end{center}
\caption{The square of the halfwidth of the absorption line profile as a function of the distance from the focal point of the focusing lens.}
\label{fig4}
\end{figure}

In order to obtain a weak  field limit for the pump (saturating) pulse
we have done a Z-scan-type measurement. In figure \ref{fig2} the lens
focuses the modulated light beam  into the sample. If the distance $z$
between the sample and the focal point of the lens is larger than the
Rayleigh-length, the  atoms in the sample experience  a field strength
proportional  to $z^{-1}$.  Hence  in Eq.~(\ref{line-shape})  the peak
depth of the spectral hole is proportional to $z^{-2}$: if we move the
sample farther and farther from  the focal point we obtain smaller and
smaller  saturation.  On  the other  hand,  the cross  section of  the
focused beam is proportional to $z^2$. Therefore, by moving the sample
along the  optical axis  the number of  atoms involved in  the process
increases quadratically,  while the signal  coming from the  atoms per
unit area decreases quadratically,  i.e. the intensity of the detected
signal does not change.   For our relatively low sensitivity detectors
this is  a great  advantage. The result  of the Z-scan  measurement is
shown in figure \ref{fig4}.  By evaluating the measured half-widths as
a   function   of   $z$   we   obtained   that   in   the   limit   of
$z\rightarrow\infty$ the half-width  of the Lorentzian-distribution in
Eq.~(\ref{line-shape}) is $\sigma=(42.7\pm0.7)  {\rm MHz}$ which yields
a dipole relaxation time of $T_2\approx90$ns.

\begin{figure}
\begin{center}
\includegraphics[width=12cm]{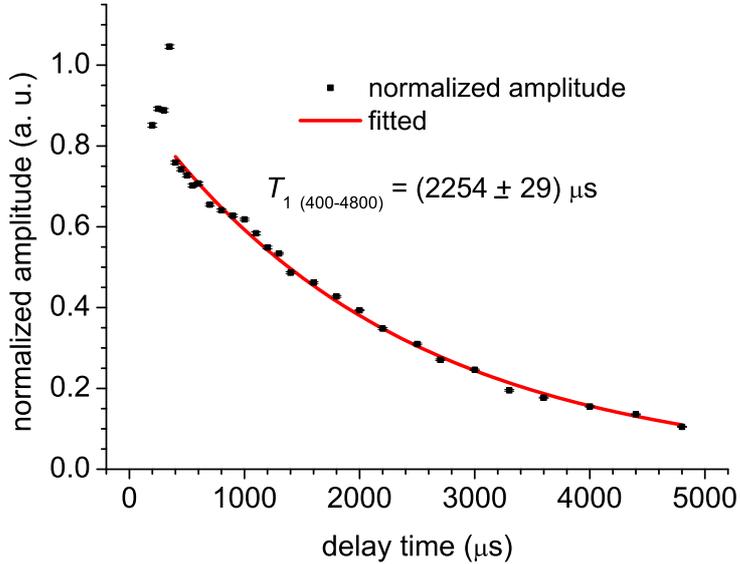}
\end{center}
\caption{The  decay  of the  peak  depth of  the  spectral  hole as  a
  function of delay time between the pump and probe pulses.}
\label{fig5}
\end{figure}

In  Eq.~(\ref{line-shape}) we  can  see  that the  peak  depth of  the
spectral hole  decays exponentially with  a rate of $T_1^{-1}$.   In a
true two-level  system this relation can  be used to  obtain the $T_1$
population relaxation  time. However, in  case of Er$^{3+}$,  there is
the  level  $^4$I$_{13/2}$   between  the  levels  $^4$I$_{11/2}$  and
$^4$I$_{15/2}$.   Hence there are  two decay  channels from  the state
$^4$I$_{11/2}$ that we excite in our pump-probe experiment. Therefore,
our model  is a rough  approximation of the real  physical situation.
Nevertheless,   the  line-shape  function   of  Eq.~(\ref{line-shape})
describes  correctly the  frequency dependence  of the  spectral hole,
because the coherent  pumping and probing do not  create any coherence
between the states  $^4$I$_{11/2}$ -- $^4$I$_{13/2}$ or $^4$I$_{15/2}$
-- $^4$I$_{13/2}$.   However, the exponential dependence  on the delay
time $t_d$ changes  in a three-level model.  In  order to estimate the
validity of  our approximations in  obtaining Eq.~(\ref{convolve}), we
performed a series of measurements in which we measured the peak depth
of the spectral  hole as a function of delay  time. The obtained curve
is shown in figure \ref{fig5}.  An exponential function fits very well
the  measured   points,  we  obtained  for  the   decay  time  constant
$T_1=(2254\pm29) \mu$s.   This justifies our  initial assumptions that
$T_2\ll T_1$.

\section{Summary}
\label{summary}

In this paper we have proposed an alternative technique to measure the
$T_2$  dipole   relaxation  time   based  on  spectral   hole  burning
spectroscopy.  Our method is applicable  if the atom can be treated as
an  effective two-level  system  interacting with  a narrow  linewidth
coherent  laser field.   It  is  shown  that  for  a  sufficiently  low
intensity  pump  and probe  fields,  the  half-width  $\sigma$ of  the
recorded  absorption line-profile  is  related to  the  $T_2$ time  as
$T_2=4/\sigma$.   Based  on the  theoretical  considerations, we  have
measured  the dipole  relaxation time  associated with  the transition
$^4$I$_{11/2}$ -- $^4$I$_{15/2}$  of erbium in LiNbO$_3$:Er$^{3+}$. We
plan to compare the measured $T_2\approx90$ns value with the result of
a photon-echo measurement.

\ack  

This work  has been  supported by  Research Fund of  the Hungarian  Academy of
Sciences (OTKA) K60086.  Z.K acknowledges the support of the Bolyai Program of
the  Hungarian  Academy  of  Sciences, and  the  Austrian-Hungarian  Bilateral
Program AT-2/2008. The  authors are grateful for the  Crystal Technology group
in our  Institute for providing the crystal.  We also thank to  G. Szab\'o and
L. M\'at\'e the technical assistance. We are indebted to P. Ma\'ak for helping
us in the AOM setup.

\end{document}